\documentclass[journal,twocolumn]{IEEEtran}
\IEEEoverridecommandlockouts
\usepackage{cite}
\usepackage{amsmath,amssymb,amsfonts,amsthm}
\DeclareMathSymbol{\shortminus}{\mathbin}{AMSa}{"39}
\usepackage{graphicx}
\usepackage{mathrsfs}

\usepackage{subfigure}
\usepackage{mathtools} 
\usepackage[linesnumbered, ruled]{algorithm2e} 
\usepackage{textcomp}
\usepackage{tabu} 
\usepackage{xcolor}
\usepackage{multirow}

\newtheorem{problem}{Problem}

\newtheorem{rem}{Remark}

\newtheorem{ass}{Assumption}

\usepackage{comment}

\newcommand{\bfx}{\mathbf{x}}
\newcommand{\bfu}{\mathbf{u}}
\newcommand{\bfy}{\mathbf{y}}
\newcommand{\bfw}{\mathbf{w}}
\newcommand{\bfv}{\mathbf{v}}
\newcommand{\N}{\mathbb{N}}
\newcommand{\R}{\mathbb{R}}
\newcommand{\E}{\mathbb{E}}
\renewcommand{\P}{\mathbb{P}}

\graphicspath{{./Figures/}}

\begin{document}

\title{Computationally Efficient Chance Constrained Covariance Control with Output Feedback\\}
\author{%
	Joshua Pilipovsky\textsuperscript{*} \thanks{\textsuperscript{*} J. Pilipovsky is a PhD student at the School of Aerospace Engineering, Georgia Institute of Technology, Atlanta, GA 30332-0150, USA. Email: jpilipovsky3@gatech.edu} 
	~~~~
	~~~~
	Panagiotis Tsiotras\textsuperscript{$\dagger$} \thanks{\textsuperscript{$\dagger$} P. Tsiotras is the David \& Lewis Chair and Professor at the School of Aerospace Engineering and the Institute for Robotics \& Intelligent Machines, Georgia Institute of Technology, Atlanta, GA 30332-0150, USA. Email: tsiotras@gatech.edu}
}%
\maketitle

\begin{abstract}
	This paper studies the problem of developing computationally efficient solutions for steering the distribution of the state of a stochastic, linear dynamical system between two boundary Gaussian distributions in the presence of chance-constraints on the state and control input.
	It is assumed that the state is only partially available through a measurement model corrupted with noise.
	The filtered state is reconstructed with a Kalman filter, the chance constraints are reformulated as difference of convex (DC) constraints, and the resulting covariance control problem is reformulated as a DC program, which is solved using successive convexification.
	The efficiency of the proposed method is illustrated on a double integrator example with varying time horizons, and is compared to other state-of-the-art chance-constrained covariance control methods.	
\end{abstract} 

\section{Introduction}
Covariance control deals with the problem of steering the entire distribution of system states from a prescribed initial distribution to a terminal distribution.
In the case the boundary conditions and noise entering the system are both normally distributed, covariance steering (CS) is equivalent to steering the first two moments of the state distribution, that is, the mean and covariance.
This fundamental problem has been studied as early back as the 1980's through the works of \cite{HS1, HS2, CovarianceAssignmentTheory}, dealing only with the infinite horizon case, where the state covariance asymptotically approaches its target value.
Of more practical interest, the finite horizon case has only been studied in recent years, starting with the works of \cite{Max1, EB1, EB2} for discrete-time dynamics and \cite{Chen, Chen2, Chen3, Halder} for continuous-time dynamics.

In practical applications, the state and control are usually subjected to certain constraints, such as the maximum thrust an engine can produce, the maximum allowable glide slope in a rocket landing, or limitations on the path of a spacecraft rendezvous maneuver.
Since the exogenous noise in the system dynamics is unbounded due to its probabilistic nature, it is difficult to enforce \textit{hard} constraints, although the novel solution in \cite{IH} uses saturation functions in the control to accomplish this task.
To this end, it is often more fruitful to enforce \textit{probabilistic} constraints on the state and control input, whereby limiting the probability of violating a certain set of constraints over the entire optimization horizon.

The chance-constrained covariance steering (CC-CS) problem has been extensively studied: References \cite{Max2, kazu_PP, CS_output_feedback_continuous, incompleteState} focus on Gaussian random variables; subsequent works have aimed in relaxing this Gaussianity assumption, with notable extensions including \cite{CS_GRF} for Gaussian random fields, \cite{CS_martingale} for martingale noise processes, \cite{nonGaussianCS} for more general, non-Gaussian distributions, and lastly \cite{DR_CS, JP_DRDS_full} solves the problem when the disturbances belong to a distributional ambiguity set.
The finite-horizon theory has also been applied to a variety of realistic scenarios, including spacecraft rendezvous \cite{JP_risk_allocation}, interplanetary trajectory optimization \cite{JoshJack}, and stochastic entry guidance \cite{stochastic_entry_guidance}, as well as to high performance, aggressive driving scenarios \cite{aggressive_driving}.

Chance constraints couple the mean and covariance dynamics together, as the constraints are functions of both the mean motion of the state and its deviation from the mean.
To handle chance constraints, \cite{kazu_PP} solves the problem by employing a convex relaxation of the terminal covariance constraint, and solves the resulting convex program as a large batch problem.
The resulting convex program becomes a quadratic semi-definite program (SDP), however, the constraints are superfluously large linear matrix inequalities (LMIs), whose optimal values are quite sparse.
Instead, \cite{exact_CS_2} solves the problem using a sequential approach, optimizing over the mean and covariance at each time step, instead of solving one large batch problem.
Instead of one large LMI for the terminal covariance constraint, the latter method solves many smaller LMIs that encode the covariance propagation, thus resulting in more computationally efficient solutions, often an order of magnitude faster \cite{exact_CS}.
This method is also more attractive from the context of \textit{data-driven} control methods, where there is no knowledge of the system matrices.
This topic has recently gained much attention due to the complexities of modeling complex systems accurately, where the theory is based on the paradigm of learning controllers directly from raw data collected from the system.
The works in \cite{DataDriven_CE, DataDriven_regularization} directly compute optimal control laws for the standard LQR problem without the intermediate step of identifying a model of the system, using similar techniques to that of \cite{exact_CS_2}.
Recently, \cite{DDCS_no_noise, DDCS_noise} has extended this data-driven framework to solve the model-free CS problem in both the settings of exact and noisy data.

The solutions to all of the above problems assume full knowledge of the state at every time step, which is often a limiting assumption on physical systems.
Reference \cite{OFCS_block} solves the CC-CS problem for the case when the state is only indirectly accessible via noise measurements, by adding a Kalman filter in the control loop.
By filtering the state and using output feedback, the control problem may be reformulated in terms of the estimated state, and subsequently solved as a convex program.
This method, as in \cite{kazu_PP}, still suffers, however, from large problem sizes due to its batch nature.

The contribution of this paper is two-fold. 
First, we solve the output feedback CC-CS (OFCC-CS) problem in a computationally efficient manner as a non-trivial extension to \cite{OFCS_block}, using the techniques employed in \cite{exact_CS_2}.
Second, we introduce a novel approach to make the state and control chance constraints tractable by reformulating them as difference of convex (DC) constraints, as opposed to linearizing the constraints about some reference values as in \cite{exact_CS_2}.
We showcase the proposed method on a double integrator path planning problem, and compare the run-time performance with that of the batch solution.

\section{Problem Statement}~\label{sec:PS}
\vspace{-0.7cm}
\subsection{Notation}~\label{subsec:notation}
Real-valued vectors are denoted by lowercase letters, $u\in\R^{m}$, matrices are denoted by uppercase letters $V\in\R^{n\times m}$, and random vectors are denoted by boldface, $\bfw\in\R^{p}$.
A random vector $\bfw$ is defined on the probability space $(\Omega, \sigma, \mathbb{P}_{\bfw})$ \cite{billingsley_probability_2012}.
We write $\sigma(\bfw)$ to denote the $\sigma$-algebra generated by the random variable $\bfw$.
We write $\bfw\sim\psi_{\bfw}$ to denote the fact that $\bfw$ is distributed according to the probability density function $\psi_{\bfw}$.
An interval of natural numbers is denoted by $\N_{[s,r]} := \{s, s+1, \ldots, r-1, r\}$, with $s, r \in \N, \ r \geq s$.
A sequence $\pi = \{u_0, \ldots, u_{N-1}\}$ is written in the short-hand notation $\{u_k\}_{k=0}^{N-1}$.
Lastly, for a symmetric matrix $\Sigma$, we write $\Sigma \succ 0 \ (\succeq 0)$ if $\Sigma$ is positive (semi-)definite.

\subsection{Problem Formulation}~\label{subsec:prob_formulation}
We consider the following discrete-time stochastic linear system
\begin{equation}~\label{eq:state_process}
	\bfx_{k+1} = A_k \bfx_k + B_k \bfu_k + G_k \bfw_k,\quad k\in\N_{[0, N-1]},
\end{equation}
where $\bfx_k\in\mathbb{R}^{n_x}$ is the state, $\bfu_k\in\mathbb{R}^{n_u}$ is the control input, $A_k\in\R^{n_x\times n_x}, B_k\in\R^{n_x\times n_u}, G_k\in\R^{n_x\times n_w}$ are the system matrices, and $N$ represents the finite optimization horizon.
The process noise $\bfw_k \sim\mathcal{N}(0,I_{n_w})$ at each time step is assumed to be i.i.d. normal random vectors.
The state is measured through the observation process
\begin{equation}~\label{eq:obs_process}
	\bfy_k = C_k \bfx_k + D_k \bfv_k, \quad k\in\N_{[0,N-1]},
\end{equation}
where $\bfy_k\in\R^{n_y}$ is the measurement and $C_k\in\R^{n_y\times n_x}, D_k\in\R^{n_y\times n_y}$ are observation matrices.
The observation noise $\bfv_k\sim\mathcal{N}(0, I_{n_y})$ at each time step is assumed to be i.i.d. normal random vectors.
\begin{ass}~\label{ass_noise}
	The observation noise matrix $D_k$ is assumed to be full-rank, hence invertible.
	The case where $D_k$ is rank-deficient can be treated using well-known techniques \cite{singular_measurements}.
\end{ass}
Before the filter is initialized, we assume some knowledge of the initial state, that is, we have an initial state estimate $\hat{\bfx}_{0^\shortminus}$ and estimation error $\tilde{\bfx}_{0^\shortminus}$, with statistics
\begin{equation}
	\hat{\bfx}_{0^\shortminus}\sim\mathcal{N}(\mu_0, \Sigma_{\hat{\bfx}_{0^\shortminus}}), \quad \tilde{\bfx}_{0^\shortminus}\sim\mathcal{N}(0, \Sigma_{\tilde{\bfx}_{0^\shortminus}}),
\end{equation}
respectively, where $\Sigma_{\hat{\bfx}_{0^\shortminus}}, \Sigma_{\tilde{\bfx}_{0^\shortminus}} \succeq 0$ and $\mu_0\in\R^{n_x}$ is known.
Using the \textit{distribution} of the initial state estimate and estimation error allows us to formulate the control law in the most general setting.
If we are given exact knowledge of the state at $k = 0$, then we set $\hat{\bfx}_{0^\shortminus} = x_0$, and $\tilde{\bfx}_{0^\shortminus} = \Sigma_{\tilde{\bfx}_{0^\shortminus}} = 0$.
Similarly, if we do not have any knowledge of the estimation error, then we set $\hat{\bfx}_{0^\shortminus} = \mu_0$ and $\Sigma_{\hat{\bfx}_{0^\shortminus}} = \tilde{\bfx}_{0^{\shortminus}} = 0$.
\begin{ass}
	We assume that the quantities $\hat{\bfx}_{0^\shortminus}, \tilde{\bfx}_{0^\shortminus}, \{\bfw_k\}_{k=0}^{N-1}$, and $\{\bfv_k\}_{k=0}^{N-1}$ are all independent.
\end{ass}

Define the filtration $\{\mathscr{F}\}_{k=-1}^{N}$ by $\mathscr{F}_{-1} = \sigma(\hat{\bfx}_{0^\shortminus})$ and $\mathscr{F}_{k} = \sigma(\hat{\bfx}_{0^\shortminus}, \bfy_i: i\in\N_{[0,k]})$ for $k\in\N_{0,N}$, which represents the information that can be used to estimate the state and control law.
For convenience, let $\mu_k := \E[\bfx_k]$ denote the mean state, and let $\hat{\bfx}_k := \E[\bfx_k | \mathscr{F}_k]$ denote the estimated state, and the estimation error as $\tilde{\bfx}_k := \bfx_k - \hat{\bfx}_k$.
Additionally, define the state, estimated state, and estimation error covariances as
\begin{subequations}~\label{eq:covariances}
	\begin{align}
		\Sigma_{\bfx_k} &= \E[(\bfx_k - \mu_k)(\bfx_k - \mu_k)^\intercal], \\
		\Sigma_{\hat{\bfx}_k} &= \E[(\hat{\bfx}_k - \mu_k)(\hat{\bfx}_k - \mu_k)^\intercal], \\
		\Sigma_{\tilde{\bfx}_k} &= \E[\tilde{\bfx}_k\tilde{\bfx}_k^\intercal] = \E[(\hat{\bfx}_k - \bfx_k)(\hat{\bfx}_k - \bfx_k)^\intercal],
	\end{align}
\end{subequations}
from which it can be shown that $\Sigma_{\bfx_k} = \Sigma_{\hat{\bfx}_k} + \Sigma_{\tilde{\bfx}_k}$.
Define the apriori estimated state and apriori estimation error as $\hat{\bfx}_{k^\shortminus} := \E[\bfx_k|\mathscr{F}_{k-1}]$ and $\tilde{\bfx}_{k^\shortminus} := \bfx_k - \hat{\bfx}_{k^\shortminus}$, respectively, with associated covariance matrices $\Sigma_{\hat{\bfx}_{k^\shortminus}}$ and $\Sigma_{\tilde{\bfx}_{k^\shortminus}}$ as given above.
To this end, it follows that the initial state is distributed as 
\begin{equation}~\label{eq:initial_state}
	\bfx_0 \sim \mathcal{N}(\mu_0, \Sigma_{\bfx_0}), \quad \Sigma_{\bfx_0} = \Sigma_{\hat{\bfx}_{0^\shortminus}} + \Sigma_{\tilde{\bfx}_{0^\shortminus}}.
\end{equation}
Additionally, we require the terminal state $\bfx_N$ to be distributed as
\begin{equation}~\label{eq:final_state}
	\bfx_{N} \sim \mathcal{N}(\mu_f, \Sigma_{\bfx_f}),
\end{equation}
or, more explicitly, that $\mu_{N} = \mu_f$ and $\Sigma_{\bfx_{N}} = \Sigma_{\hat{\bfx}_{N}} + \Sigma_{\tilde{\bfx}_{N}} = \Sigma_{\bfx_f}$.
We assume a quadratic objective function given by
\begin{equation}~\label{eq:cost}
	\mathcal{J}(\bfu) = \E\left[\sum_{k=0}^{N-1} \left((\bfx_k - x_k^{(r)})^\intercal Q_k (\bfx_k - x_k^{(r)}) + \bfu_k^\intercal R_k \bfu_k\right)\right],
\end{equation}
for a given sequence of state weight matrices $\{Q_k\}_{k=0}^{N-1} \succeq 0$ and control weight matrices $\{R_k\}_{k=0}^{N-1} \succ 0$, respectively, where $\{x_k^{(r)}\}_{k=0}^{N}$ denotes a reference state trajectory.

Lastly, we assume that the control input $\bfu_k$ is an affine function of the measurement data, and define a control sequence $\{\bfu_k\}_{k=0}^{N-1}$ to be \textit{admissible} if it satisfies this property over the entire horizon.
This requirement is made to ensure that the distribution of the state, and as a result the estimated state, will be Gaussian over the time horizon, since as the initial state $\bfx_0$ is Gaussian and the system dynamics and measurements are linear in the state and input, then the state will remain normally distributed.
The stochastic optimal control problem is now stated below.
\begin{problem}~\label{problem:CS_general}
	For a given initial state distribution \eqref{eq:initial_state}, find the admissible control sequence $\pi := \{\bfu_k\}_{k=0}^{N-1}$ that minimizes the cost functional \eqref{eq:cost} subject to the dynamics \eqref{eq:state_process} and measurement model \eqref{eq:obs_process}, such that the terminal state satisfies \eqref{eq:final_state}.
\end{problem}
	

\section{Kalman Filter}~\label{sec:KF}
For a normally distributed state that follows linear dynamics and has a linear measurement model, the optimal observer $\{L_k\}_{k=0}^{N-1}$ that minimizes the $\mathcal{H}_{2}$ norm of the estimation error $\{\tilde{\bfx}_k\}_{k=0}^{N}$ over the time horizon is given by the Kalman filter\cite{KF}. 
In the discrete-time setting for given step $k$, the filter updates the estimated state as follows.
First, a time update is made by propagating the previous estimated state forward in time using the mean dynamics as
\begin{equation}~\label{eq:time_update_state}
	\hat{\bfx}_{k^\shortminus} = A_{k-1} \hat{\bfx}_{k-1} + B_{k-1} \bfu_{k-1}.
\end{equation}
Then, the a priori estimated state at the current time step is corrected using the current measurement as 
\begin{equation}~\label{eq:measurement_update_state}
	\hat{\bfx}_{k} = \hat{\bfx}_{k^\shortminus} + L_k(\bfy_k - C_k \hat{\bfx}_{k^\shortminus}),
\end{equation}
where $L_k$ is the Kalman gain given by
\begin{equation}~\label{eq:Kalman_gain}
	L_k = \Sigma_{\tilde{\bfx}_{k^\shortminus}}C_k^\intercal(C_k \Sigma_{\tilde{\bfx}_{k^\shortminus}}C_k^\intercal + D_k D_k^\intercal)^{-1},
\end{equation}
and the state estimation error covariance are similarly updated according to the time and measurement updates
\begin{align}
	\Sigma_{\tilde{\bfx}_{k^\shortminus}} &= A_{k-1} \Sigma_{\tilde{\bfx}_{k-1}} A_{k-1}^\intercal + G_{k-1} G_{k-1}^\intercal, \label{eq:time_update_cov} \\
	\Sigma_{\tilde{\bfx}_k} &= (I_{n_x} - L_k C_k)\Sigma_{\tilde{\bfx}_{k^\shortminus}}(I_{n_x} - L_k C_k)^\intercal + L_k D_k D_k^\intercal L_k^\intercal. \label{eq:measurement_update_cov}
\end{align}
\begin{rem}
	The state error process covariance is completely independent of the control law used, and only a function of the observation model.
	Thus, it may be pre-computed before the control optimization begins.
\end{rem}
Following the discussion in \cite{OFCS_block}, the estimated state process may be rewritten as
\begin{equation}~\label{eq:estimated_state_process}
	\hat{\bfx}_{k+1} = A_k \hat{\bfx}_k + B_k \bfu_k + L_{k+1}\tilde{\bfy}_{(k+1)^\shortminus},
\end{equation}
where $\hat{\bfx}_0 = \hat{\bfx}_{0^\shortminus} + L_0\tilde{\bfy}_{0^\shortminus}$ and $\{\tilde{\bfy}_{k^\shortminus}\}_{k=0}^{N}$ is defined as the \textit{innovation} process
\begin{equation}~\label{eq:innovation_process}
	\tilde{\bfy}_{k^\shortminus} := \bfy_k - \E[\bfy_k | \mathscr{F}_{k-1}], \quad k\in\N_{[0, N]},
\end{equation}
which is distributed as $\tilde{\bfy}_{k^\shortminus} \sim \mathcal{N}(0, \Sigma_{\tilde{\bfy}_{k^\shortminus}})$, where the covariance of the innovation is
\begin{equation}~\label{eq:innovation_cov}
	\Sigma_{\tilde{\bfy}_{k^\shortminus}} := \E[\tilde{\bfy}_{k^\shortminus}\tilde{\bfy}_{k^\shortminus}^\intercal] = C_k \Sigma_{\tilde{\bfx}_{k^\shortminus}} C_k^\intercal + D_k D_k^\intercal.
\end{equation}
Returning to \eqref{eq:estimated_state_process}, we have essentially converted the state and observation process models, \eqref{eq:state_process} and \eqref{eq:obs_process}, respectively, with a corresponding filtered state process with noise $L_{k+1}\tilde{\bfy}_{(k+1)^\shortminus}$.
Lastly, minimizing the cost \eqref{eq:cost} is equivalent to minimizing the following reformulated cost in terms of the estimated state
\begin{equation}~\label{eq:cost_filtered_state}
	\mathcal{J} = \E\left[\sum_{k=0}^{N-1} (\hat{\bfx}_k^\intercal Q_k \hat{\bfx}_k + \bfu_k^\intercal R_k \bfu_k)\right] - 2\sum_{k=0}^{N-1}\mu_k^\intercal Q_k x_k^{(r)}.
\end{equation}
See \cite{OFCS_block} for details on how to compute this equivalent cost and the innovation process covariance.
In terms of the estimated state, the terminal constraints may be written as
\begin{equation}~\label{eq:terminal_filtered_state}
	\hat{\bfx}_{N} \sim \mathcal{N}(\mu_{f}, \Sigma_{\bfx_f} - \Sigma_{\tilde{\bfx}_{N}}),
\end{equation}
from which we see that in order for $\Sigma_{\bfx_f} = \Sigma_{\bfx_N}$, it follows that $\Sigma_{\bfx_f} \succ \Sigma_{\tilde{\bfx}_{N}}$.
The covariance steering problem is now posed as follows.
\begin{problem}~\label{problem:CS_filtered_state}
	Find the admissible control sequence $\pi := \{\bfu_k\}_{k=0}^{N-1}$ that minimizes the cost functional \eqref{eq:cost_filtered_state} subject to the filtered state dynamics \eqref{eq:estimated_state_process} and terminal constraint \eqref{eq:terminal_filtered_state} for a given $\Sigma_{\bfx_f} \succ \Sigma_{\tilde{\bfx}_{N}}$.
\end{problem}


\section{Filtered State Control Design}~\label{sec:control_design}
We consider an affine filtered state feedback control design of the form
\begin{equation}~\label{eq:control_law}
	\bfu_k = K_k (\hat{\bfx}_k - \mu_k) + m_k, \quad k\in\N_{[0,N-1]},
\end{equation}
where $K_k\in\R^{m\times n}$ is the feedback gain matrix that controls the covariance of the filtered state, and $m_k\in\R^{m}$ is the feed-forward term that controls the state mean.
Using this control structure, the objective function \eqref{eq:cost_filtered_state} may be rewritten as
\begin{align}
	\mathcal{J}(K_k, m_k) &= \sum_{k=0}^{N-1} \Big(\mathrm{tr}(Q_k\Sigma_{\hat{\bfx}_{k}}) + \mathrm{tr}(R_k K_k \Sigma_{\hat{\bfx}_k} K_k^\intercal) \nonumber \\
	&+ \mu_k^\intercal Q_k \mu_k + m_k^\intercal R_k m_k - 2\mu_k^\intercal Q_k x_k^{(r)}\Big),
\end{align}
which, equivalently, may be expressed as
\begin{subequations}
	\begin{align}
		\mathcal{J}(m_k, K_k) &= \mathcal{J}_{\mu}(m_k; \mu_k) + \mathcal{J}_{\Sigma}(K_k; \Sigma_{\hat{\bfx}_k}), \nonumber \\
		\mathcal{J}_{\mu}(m_k; \mu_k) &= \sum_{k=0}^{N-1} \left(\mu_k^\intercal Q_k \mu_k + m_k^\intercal R_k m_k - 2\mu_k^\intercal Q_k x_k^{(r)}\right), \label{eq:mean_cost} \\
		\mathcal{J}_{\Sigma}(K_k; \Sigma_{\hat{\bfx}_k}) &= \sum_{k=0}^{N-1} \left(\mathrm{tr}(Q_k \Sigma_{\hat{\bfx}_k}) + \mathrm{tr}(R_k K_k \Sigma_{\hat{\bfx}_k}K_k^\intercal)\right), \label{eq:cov_cost}
	\end{align}
\end{subequations}
that is, separable in terms of a mean cost $\mathcal{J}_{\mu}$ and a covariance cost $\mathcal{J}_{\Sigma}$.
Plugging the control law \eqref{eq:control_law} into the filtered state dynamics \eqref{eq:estimated_state_process} and computing the first two moments yields the following mean and covariance dynamics of the filtered state as follows
\begin{subequations}~\label{eq:filtered_state_moment_dynamics}
	\begin{align}
		\mu_{k+1} &= A_k \mu_k + B_k m_k, \label{eq:mean_dynamics}\\
		\Sigma_{\hat{\bfx}_{k+1}} &= (A_k + B_k K_k) \Sigma_{\hat{\bfx}_{k}} (A_k + B_k K_k)^\intercal \nonumber \\
		&\hspace{3cm} + L_{k+1}\Sigma_{\tilde{\bfy}_{(k+1)^\shortminus}} L_{k+1}^\intercal. \label{eq:cov_dynamics}
	\end{align}
\end{subequations}
Problem~\ref{problem:CS_filtered_state} is then equivalent to the following two sub-problems, the first for the mean control, and the second for the covariance control.
\begin{problem}~\label{problem:mean_steering}
	Find the sequence of mean controls $\{m_k\}_{k=0}^{N-1}$ that solves the mean steering problem
	\begin{align*}
		&\min_{m_k, \mu_k} \ \eqref{eq:mean_cost} \\
		&\quad \mathrm{s.t.} \quad \eqref{eq:mean_dynamics}, \quad k\in\N_{[0, N-1]}, \\
		&\quad\quad\quad \ \mu_N = \mu_f.
	\end{align*}
\end{problem}
\begin{problem}~\label{problem:cov_steering}
	Find the sequence of feedback gains $\{K_k\}_{k=0}^{N-1}$ that solves the covariance steering problem
	\begin{align*}
		&\min_{K_k, \Sigma_{\hat{\bfx}_k}} \ \eqref{eq:cov_cost} \\
		&\quad \mathrm{s.t.} \quad \eqref{eq:cov_dynamics}, \quad k\in\N_{[0,N-1]}, \\
		&\quad\quad\quad \ \Sigma_{\hat{\bfx}_{N}} = \Sigma_{\bfx_f} - \Sigma_{\tilde{\bfx}_{N}}.
	\end{align*}
\end{problem}
The mean steering problem is a standard convex quadratic program, and it is straightforward to compute the solution analytically, thus we omit the discussion for this.
The covariance steering problem, however, is non-convex, due to both the covariance dynamics \eqref{eq:cov_dynamics} and the nonlinear cost \eqref{eq:cov_cost}, since the filtered state covariance is part of the decision variables.
To this end, by defining the new decision variables $U_k := K_k \Sigma_{\hat{\bfx}_k}$, Problem~\ref{problem:cov_steering} is equivalently written as
\begin{subequations}~\label{eq:cov_steering_change_of_variables}
	\begin{align}
		&\min_{U_k, \Sigma_{\hat{\bfx}_k}} \ \mathcal{J}_{\Sigma} = \sum_{k=0}^{N-1} \left(\mathrm{tr}(Q_k \Sigma_{\hat{\bfx}_k}) + \mathrm{tr}(R_k U_k \Sigma_{\hat{\bfx}_k}^{-1} U_k^\intercal)\right) \\
		&\quad \mathrm{s.t.} \quad A_k\Sigma_{\hat{\bfx}_k}A_k^\intercal + B_k U_k A_k^\intercal + A_k U_k^\intercal B_k^\intercal \nonumber \\
		&\quad + L_{k+1}\Sigma_{\tilde{\bfy}_{(k+1)^\shortminus}} L_{k+1}^\intercal + B_k U_k \Sigma_{\hat{\bfx}_{k}}^{-1}U_k^\intercal B_k^\intercal - \Sigma_{k+1} = 0, \\
		&\quad\quad\quad \ \Sigma_{\hat{\bfx}_{N}} = \Sigma_{\bfx_f} - \Sigma_{\tilde{\bfx}_{N}}.
	\end{align}
\end{subequations}
The optimization problem \eqref{eq:cov_steering_change_of_variables} is still non-convex due to the presence of the $U_k\Sigma_{\hat{\bfx}_{k}}^{-1} U^\intercal_k$ terms in the cost and dynamics.
To this end, define $Y_k := U_k\Sigma_{\hat{\bfx}_{k}}^{-1} U^\intercal_k$ and relax the problem to
\begin{subequations}~\label{eq:cov_steering_change_of_variables_relax}
	\begin{align}
		&\min_{Y_k, U_k, \Sigma_{\hat{\bfx}_k}} \ \mathcal{J}_{\Sigma} = \sum_{k=0}^{N-1} \left(\mathrm{tr}(Q_k \Sigma_{\hat{\bfx}_k}) + \mathrm{tr}(R_k Y_k)\right) \\
		&\quad \ \mathrm{s.t.} \quad \forall k\in\N_{[0, N-1]}, \nonumber \\
		&\quad\quad\quad \ C_k := U_k\Sigma_{\hat{\bfx}_{k}}^{-1} U^\intercal_k - Y_k \preceq 0, \label{eq:relaxation_LMI} \\
		&\quad\quad\quad \ G_k^{(1)} := A_k\Sigma_{\hat{\bfx}_k}A_k^\intercal + B_k U_k A_k^\intercal + A_k U_k^\intercal B_k^\intercal \nonumber \\
		&\quad + L_{k+1}\Sigma_{\tilde{\bfy}_{(k+1)^\shortminus}} L_{k+1}^\intercal + B_k Y_k B_k^\intercal - \Sigma_{\hat{\bfx}_{k+1}} = 0, \\
		&\quad\quad\quad \ G_k^{(2)} := -\Sigma_{\hat{\bfx}_{N}} + \Sigma_{\bfx_f} - \Sigma_{\tilde{\bfx}_{N}} = 0.
	\end{align}
\end{subequations}
The optimization problem \eqref{eq:cov_steering_change_of_variables_relax} is now convex, since the constraint \eqref{eq:relaxation_LMI} can be written using the Schur complement as the linear matrix inequality (LMI)
\begin{equation}
	\begin{bmatrix}
		\Sigma_{\hat{\bfx}_{k}} & U_k^\intercal \\
		U_k & Y_k
	\end{bmatrix} \succeq 0.
\end{equation}
Thus, the resulting problem becomes a semi-definite program (SDP), which is convex and thus has a global minimum.
The optimal feedback gains may be recovered from the decision variables as $K_k = U_k \Sigma_{\hat{\bfx}_{k}}^{-1}$.
It turns out the relaxation imposed to get \eqref{eq:cov_steering_change_of_variables_relax} is a \textit{lossless} relaxation, which implies that the solution to the relaxed problem \eqref{eq:cov_steering_change_of_variables} is equivalent to the solution to the original problem \eqref{eq:cov_steering_change_of_variables_relax}, or more explicitly that $C_k \equiv 0, \ \forall k\in\N_{[0,N-1]}$.
See \cite{exact_CS} for the proof in the full-state feedback case, which naturally extends to the present case.


\section{Chance Constraints}
In real-world applications, there is usually some kind of domain $\mathcal{X}\subset\R^{n}, \mathcal{U}\subset\R^{m}$ that the state and control must reside in, respectively.
These domains represent the physical limitations of the system, for example maximum propulsive thrust of a rocket motor \cite{JoshJack} or smoothly landing in a conical state space \cite{PDG_Jack}.
Due to the nature of the dynamics model \eqref{eq:state_process}, the exogenous noise entering the system is \textit{unbounded}, so it is difficult to impose exact constraints on the state and control input.
To remedy this, we impose \textit{probabilistic}, or chance constraints, on the state and control, limiting the joint probability of violating the constraints over the entire time horizon.

For simplicity, we assume that the constraint spaces for the state and input are modeled as polytopes, that is,
\begin{subequations}~\label{eq:polytopes}
	\begin{align}
		\mathcal{X}_k &:= \cap_{i=1}^{N_c^x} \{x : \alpha_{k, i}^\intercal x \leq \beta_{k, i} \}, \quad k\in\N_{[1, N]} \subset \R^{n}, \label{eq:state_polytope} \\
		\mathcal{U}_k &:= \cap_{i=1}^{N_c^u} \{u : a_{k, i}^\intercal u \leq b_{k, i} \}, \quad k\in\N_{[0, N - 1]} \subset \R^{m}, \label{eq:input_polytope}
	\end{align}
\end{subequations}
where $\alpha_{k, i}, a_{k, i}\in\R^{n}$ and $\beta_{k,i}, b_{k, i}\in\R$.
The following discussion may also be applied to convex cone constraint sets using the techniques outlined in \cite{JP_risk_allocation}.
We require that the joint chance constraint over the time horizon is less than a pre-specified threshold, i.e.,
\begin{subequations}~\label{eq:jointCC}
	\begin{align}
		&\P\left(\bigwedge_{k=1}^{N} \bfx_k \notin \mathcal{X}_k\right)\leq \Delta_{x}, \label{eq:state_jointCC} \\
		&\P\left(\bigwedge_{k=0}^{N-1} \bfu_k \notin \mathcal{U}_k\right)\leq \Delta_{u}, \label{eq:input_jointCC}
	\end{align}
\end{subequations}
where $\P(\cdot)$ denotes the probability of an event, and $\Delta_x, \Delta_u \in (0, 0.5]$.
Given the polytope constraint spaces \eqref{eq:polytopes}, the chance constraints \eqref{eq:jointCC} can be written equivalently as 
\begin{subequations}~\label{eq:polytope_jointCC}
	\begin{align}
		&\P\left(\bigwedge_{k=1}^{N}\bigwedge_{i=1}^{N_c^x}\alpha_{k,i}^\intercal \bfx_k > \beta_{i, k}\right) \leq \Delta_{x}, \label{eq:polytope_joint_CC_state} \\
		&\P\left(\bigwedge_{k=0}^{N-1}\bigwedge_{i=1}^{N_c^u} a_{k,i}^\intercal \bfu_k > b_{i, k}\right) \leq \Delta_{u}, \label{eq:polytope_joint_CC_input}
	\end{align}
\end{subequations}
Using Boole's inequality \cite{Boole}, one can conservatively decompose a \textit{joint} chance constraint to the individual chance constraints
\begin{subequations}~\label{eq:polytope_individualCC}
	\begin{align}
		&\P(\alpha_{i, k}^\intercal \bfx_k \leq \beta_{k, i}) \geq 1 - \delta_{i, k}^{x}, \quad k\in\N_{[1, N]}, \ i\in\N_{[1, N_c^x]}, \label{eq:polytope_individualCC_state} \\
		&\P(a_{i, k}^\intercal \bfu_k \leq b_{k, i}) \geq 1 - \delta_{i, k}^{u}, \quad k\in\N_{[0, N - 1]}, \ i\in\N_{[1, N_c^u]}, \label{eq:polytope_individualCC_input}
	\end{align}
\end{subequations}
with
\begin{equation}~\label{eq:risk_allocations}
	\sum_{k=1}^{N} \sum_{i=1}^{N_c^x} \delta_{i, k}^{x} \leq \Delta_{x}, \quad \sum_{k=0}^{N - 1} \sum_{i=1}^{N_c^u} \delta_{i, k}^{u} \leq \Delta_{u},
\end{equation}
where $\delta_{i,k}\in [0, \Delta]$ represents the probability of violating the $i$th constraint at time step $k$.
For simplicity, in this work we will assume that the risk allocations $\{\delta_{i,k}\}$ are uniformly allocated and constant, i.e., $\delta_{i,k} = \delta, \ \forall i, k$, where $\delta = \Delta / (N N_c)$.
\begin{rem}
	By fixing the risk allocations, the resulting optimization problem is sub-optimal since the risks themselves are decision variables.
	However, bi-level methods \cite{JP_risk_allocation} have been developed that jointly optimize the risk as well as the control law while guaranteeing local optimality.
\end{rem}
It can be shown that in the case the state and input are normal random variables, the chance constraints \eqref{eq:polytope_individualCC} may be re-written in terms of the first two moments
\begin{subequations}~\label{eq:polytope_individualCC_moments}
	\begin{align}
		&\Phi^{-1}(1 - \delta_{i,k}^{x})\sqrt{\alpha_{i,k}^\intercal \Sigma_{\bfx_k} \alpha_{i,k}} + \alpha_{i,k}^\intercal \mu_k \leq \beta_{i,k}, \label{eq:polytope_individualCC_state_moments} \\
		&\Phi^{-1}(1 - \delta_{i,k}^{u})\sqrt{a_{i,k}^\intercal Y_k a_{i,k}} + a_{i,k}^\intercal m_k \leq b_{i,k}, \label{eq:polytope_individualCC_input_moments} 
	\end{align}
\end{subequations}
where $\Phi^{-1}(\cdot)$ is the inverse standard normal cumulative distribution function.
If the state and input are not Gaussian, one may use various concentration inequalities \cite{DR_CS} to conservatively approximate the chance constraints.
Note that neither the state constraint \eqref{eq:polytope_individualCC_state_moments} nor the control constraint \eqref{eq:polytope_individualCC_input_moments} are convex due to the square root of the decision variables $\Sigma_{k}$ and $Y_{k}$, respectively.

In \cite{exact_CS_2}, the authors approximate the chance constraints by linearizing around some reference values $\Sigma_{r}$ and $Y_{r}$, respectively.
This leads to a tractable convex program albeit a conservative one.
The details of the convexification are shown in Appendix~A.
Here, instead, we notice that the chance constraints may be written as a difference of convex (DC) functions.
To this end, we can equivalently write the constraints \eqref{eq:polytope_individualCC_moments} by squaring both sides sides as
\begin{subequations}
	\begin{align}
		&(\Phi^{-1}(1 - \delta_{i,k}^{x}))^{2}\alpha_{i,k}^\intercal(\Sigma_{\hat{\bfx}_k} + \Sigma_{\tilde{\bfx}_k})\alpha_{i,k}^\intercal \leq (\beta_{i,k} - \alpha_{i,k}^\intercal\mu_k)^2, \label{eq:squared_CC_state1} \\
		&\beta_{i,k} - \alpha_{i,k}^\intercal \mu_k \geq 0, \label{eq:squared_CC_state2} \\
		&(\Phi^{-1}(1 - \delta_{i,k}^{u}))^{2}a_{i,k}^\intercal Y_k a_{i,k} \leq (b_{i,k} - a_{i,k}^\intercal m_k)^2, \label{eq:squared_CC_input1} \\
		&b_{i,k} - a_{i,k}^\intercal m_k \geq 0, \label{eq:squared_CC_input2}
	\end{align}
\end{subequations}
where the constraints \eqref{eq:squared_CC_state2} and \eqref{eq:squared_CC_input2} are needed to enforce the equivalence with \eqref{eq:polytope_individualCC_moments}.
Further, \eqref{eq:squared_CC_state1} and \eqref{eq:squared_CC_input1} may be written as
\begin{subequations}
	\begin{align}
		f_{x}(\Sigma_{\hat{\bfx}_k}; \alpha_{i,k}, \delta_{i,k}^{x}, \Sigma_{\tilde{\bfx}_k}) - g_{x}(\mu_{k}; \alpha_{i,k}, \beta_{i,k}) &\leq 0, \label{eq:DC_state} \\
		f_{u}(Y_k; a_{i,k}, \delta_{i,k}^{u}) - g_{u}(m_k; a_{i,k}, b_{i,k}) &\leq 0, \label{eq:DC_input}
	\end{align}
\end{subequations}
where $f$ and $g$ are both convex in the decision variables.
DC constraints may be handled by using the convex-concave procedure (CCP) \cite{CCP_DC_programming}, which is guaranteed to converge to a feasible point.
See Appendix~B for details on the CCP in the context of the chance constraints \eqref{eq:polytope_individualCC_state_moments}, \eqref{eq:polytope_individualCC_input_moments}.


\section{Numerical Example}~\label{sec:examples}
We illustrate the output feedback covariance steering (OFCS) algorithm on a double integrator system with horizon $N = 20$ and $\Delta t = 0.2$.
The system dynamics are given by
\begin{equation}
	A = 
	\begin{bmatrix}
		1 & 0 & \Delta t & 0 \\
		0 & 1 & 0 & \Delta t \\
		0 & 0 & 1 & 0 \\
		0 & 0 & 0 & 1
	\end{bmatrix}, \quad
	B = 
	\begin{bmatrix}
		\Delta t^2 / 2 & 0 \\
		0 & \Delta t^2 / 2 \\
		\Delta t & 0 \\
		0 & \Delta t
	\end{bmatrix},
\end{equation}
with process noise matrix $G_k = 0.01 \times I_{3}$.
The measurement model is given by $C_k = [0_{3\times 1}, I_{3}]$, meaning we are measuring the last three states, and measurement noise matrix $D_k = \mathrm{diag}(0.1, 0.003, 0.003)$.
The initial state estimation and error are given by $\Sigma_{\hat{\bfx}_{0^\shortminus}} = \mathrm{diag}(8, 9, 0.6, 0.6)\times 10^{-2}$ and $\Sigma_{\tilde{\bfx}_{0^\shortminus}} = \mathrm{diag}(2, 1, 1.4, 1.4)\times 10^{-2}$, respectively, and the initial state mean is $\mu_0 = [1.5, 3.5, 1.5, 8.5]^\intercal$.
The desired terminal state distribution has mean $\mu_{f} = [10.5, 8.5, 0, 0]^\intercal$ and covariance $\Sigma_{\bfx_f} = \mathrm{diag}(3, 3, 0.3, 0.3)\times 10^{-2}$.
Lastly, we enforce state chance constraints where the polytope $\mathcal{X}$ is defined for $N_c^{x} = 2$ half spaces given by $\alpha_1 = [-0.25, 1, 0, 0]^\intercal, \ \alpha_2 = [0.5, -1, 0, 0]^\intercal$, and $\beta_1 = 6, \ \beta = -2.1$, with a joint risk level $\Delta_{x} = 0.02$.
We solve the resulting optimization problem with the CCP using YALMIP \cite{YALMIP} with MOSEK \cite{MOSEK}.

\begin{figure}[!htb]
	\centering
	\subfigure[Optimal $3\sigma$ covariance ellipses for OFCS with no constraints.]{
		\includegraphics[scale=0.345]{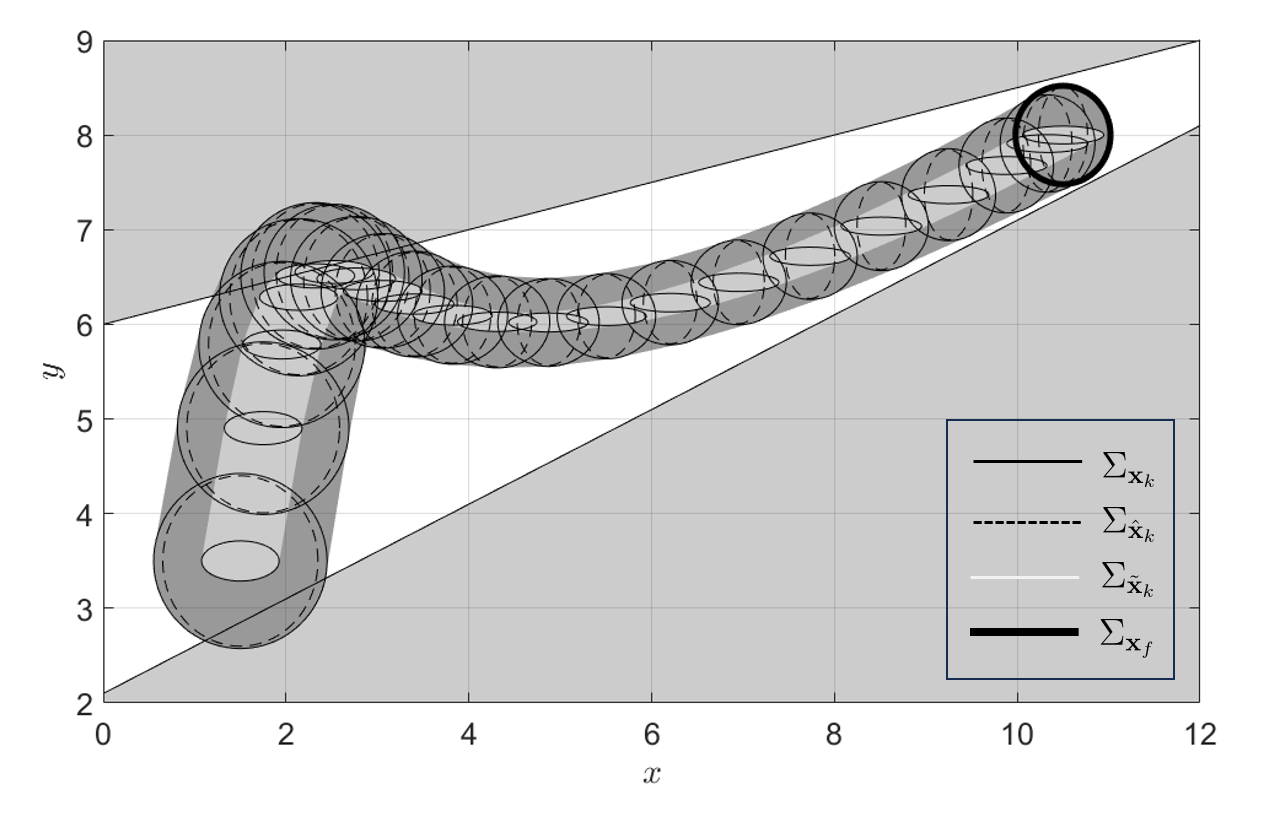}
		\label{fig:trajs_no_constraints}
	}
	\vspace{1em} 
	\subfigure[Optimal $3\sigma$ covariance ellipses for OFCS with chance constraints.]{
		\includegraphics[scale=0.35]{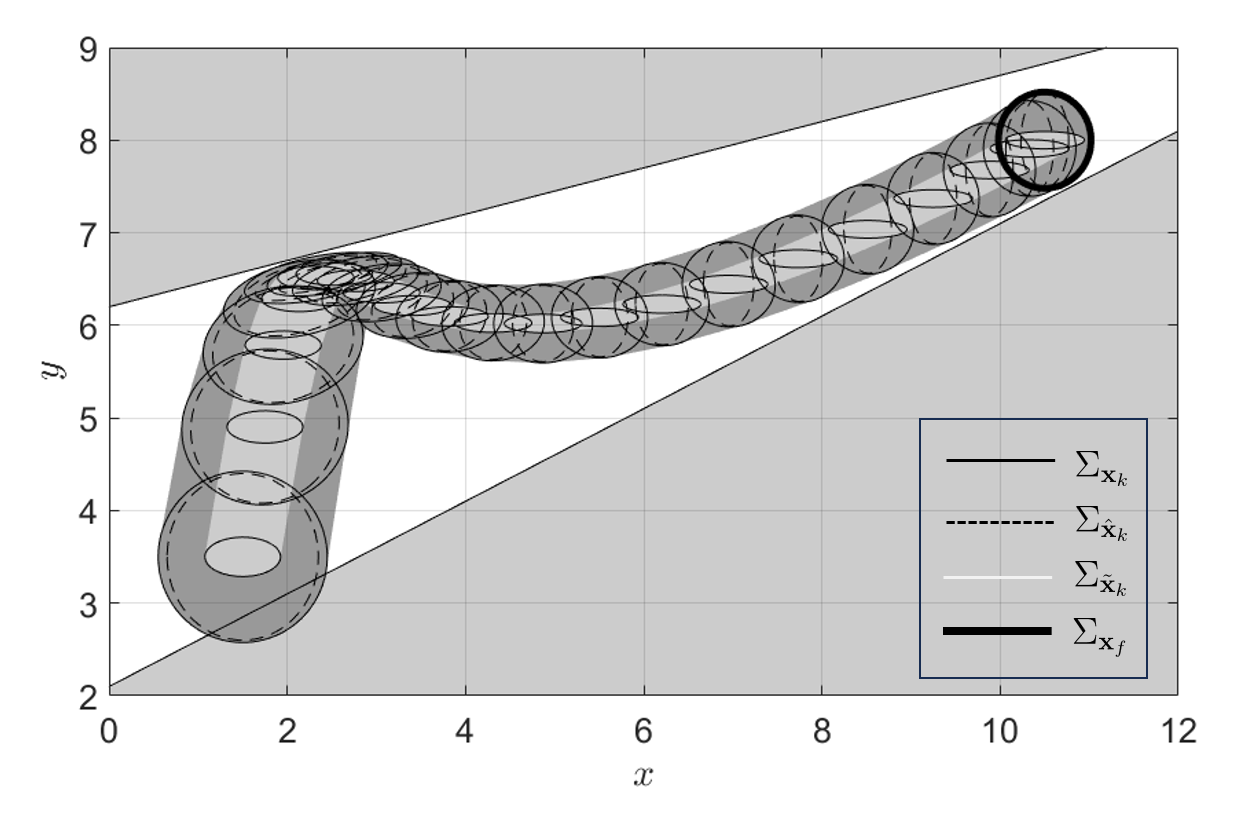}
		\label{fig:trajs_chance_constraints}
	}
	\caption{Evolution of 3$\sigma$ covariance ellipses for the true state, estimated state, and state estimation error.}
	\label{fig:optimal_trajectories}
\end{figure}

Figure~\ref{fig:optimal_trajectories} shows the evolution of the optimal 3-$\sigma$ covariances.
We see that indeed the chance constraints are satisfied, and the terminal covariance constraint is met with equality.
We also adjust the time horizon $N$ and compare the run times to that of the approach in \cite{OFCS_block}.

\renewcommand*{\arraystretch}{1.5}
\begin{table}[!htb]
	\centering
	\caption{Statistics of solution times for varying horizon lengths.}
	\begin{tabular}{lllll}
		\hline
		$N$ & 10 & 20 & 50 & 100 \\
		\hline
		\multicolumn{5}{c}{Block CS} \\
		\hline
		$\mu$ (ms) & 25.17 & 71.14 & 535.44 & 3827.02 \\
		$\sigma$ (ms) & 1.84 & 1.50 & 8.28 & 38.67 \\
		\hline
		\multicolumn{5}{c}{Sequential CS} \\
		\hline
		$\mu$ (ms) & 6.43 & 10.43 & 27.52 & 47.83 \\
		$\sigma$ (ms) & 0.52 & 0.68 & 0.72 & 1.57 \\
		\hline
	\end{tabular}
\end{table}
The proposed approach performs about an order of magnitude faster than the batch approach, and scales better with larger problem sizes.
We also see that this method has more consistent performance, as the standard deviations of the solution run times are smaller.
Further, we see that the batch solution scales quadratically with the problem size, however our method scales linearly.
Specifically, for a large horizon length $N = 100$ (20 seconds), the proposed solution is almost a hundred times faster than the batch solution.
As noted in \cite{exact_CS_2}, the terminal covariance constraints for the SDP in \eqref{eq:cov_steering_change_of_variables_relax} require only $N - 1$ LMIs of dimension $m\times m$, whereas the approach in \cite{JoshJack} requires an LMI of dimension $(N+2)n \times (N+2)n$.
In general, many smaller LMIs are more efficiently solved compared to that of a single large one.


\section{Conclusion}~\label{sec:conclusion}
We have extended the recently developed computationally efficient approach to solve the CC-CS problem to problems with partial state information.
The OFCC-CS problem is posed as a convex program by first filtering the state using a Kalman filter, and subsequently solving a CC-CS problem for the estimated state evolution.
We also introduce a new approach to handle chance constraints by formulating them as DC constraints and successively convexifying them using the convex-concave procedure, leading to less conservative optimal trajectories.
Future work will aim to extend the approach described in this work to a data-driven context, where the system dynamics are assumed to be unknown, and only collected data through experiment is used to perform control design and synthesis.


\section{Acknowledgment}

This work has been supported by NASA University Leadership Initiative award 80NSSC20M0163 and ONR award N00014-18-1-2828.
The article solely reflects the opinions and conclusions of its authors and not any NASA entity.


\bibliography{refs.bib}
\bibliographystyle{IEEEtran}


\appendix

\section*{A.~Chance Constraint Linearization}

\setcounter{equation}{0}
\renewcommand{\theequation}{A.\arabic{equation}}
The chance constraints \eqref{eq:polytope_individualCC_state_moments} and \eqref{eq:polytope_individualCC_input_moments} are non-convex in the decision variables $\Sigma_{\bfx_k}$ and $Y_{k}$ due to the square root terms.
Since the square root is a concave function, the tangent line can serve as linear global overestimator.
To this end, the authors in \cite{exact_CS_2} linearize the constraints about some reference values $\{\Sigma_r, Y_r\}$.
In practice, one can first solve a non chance-constrained CS problem to yield an initial $\{\Sigma_k\}_{k=1}^{N}, \{Y_k\}_{k=1}^{N}$ trajectory, which provides the first batch of linearization points for the chance-constraints.
The linearized chance-constraints have the form
\begin{subequations}
	\begin{align}
		\ell_{i,k}^\intercal (\Sigma_{\hat{\bfx}_k} + \Sigma_{\tilde{\bfx}_k}) \ell_{i,k} + \alpha_{i,k}^\intercal \mu_k &\leq \beta_{i,k} - \gamma_{i,k}, \\
		e_{i,k}^\intercal Y_k e_{i,k} + a_{i,k}^\intercal m_k &\leq b_{i,k} - \varphi_{i,k},
	\end{align}
\end{subequations}
where the constants $\ell, \gamma, e, \varphi$ are given by
\begin{subequations}
	\begin{align*}
		\ell_{i,k} &= \frac{\Phi^{-1}(1 - \delta_{i,k}^{x})}{\sqrt{2\alpha_{i,k}^\intercal \Sigma_{\bfx_k,r} \alpha_{i,k}}}\alpha_{i,k}, \\
		\gamma_{i,k} &= \frac{\Phi^{-1}(1 - \delta_{i,k}^{x})}{2}\sqrt{\alpha_{i,k}^\intercal \Sigma_{\bfx_{k,r}} \alpha_{i,k}}, \\
		e_{i,k} &= \frac{\Phi^{-1}(1 - \delta_{i,k}^{u})}{\sqrt{2a_{i,k}^\intercal Y_{k, r} a_{i,k}}}a_{i,k}, \\
		\varphi_{i,k} &= \frac{\Phi^{-1}(1 - \delta_{i,k}^{u})}{2}\sqrt{a_{i,k}^\intercal Y_{k, r} a_{i,k}}.
	\end{align*}
\end{subequations}
\section*{B.~Difference of Convex Programming}

\setcounter{equation}{0}
\renewcommand{\theequation}{B.\arabic{equation}}
DC programming problems have the form
\begin{subequations}~\label{eq:DC_program}
	\begin{align}
		&\min_{x\in\R^{n}} \ f_0(x) - g_0(x), \\
		&\ \mathrm{s.t.} \quad f_i(x) - g_i(x) \leq 0, \quad i = 1,\ldots, M,
	\end{align}
\end{subequations}
where $x$ is the decision variable and $f_i:\R^{n}\rightarrow\R$ and $g_i:\R^{n}\rightarrow\R, \ i=0,\ldots,M$ are convex functions.
This problem is, in general, not convex \cite{CCP_Boyd} unless the functions $g_i$ are affine.
One method to find tractable solutions to \eqref{eq:DC_program} is the convex-concave procedure (CCP) which is a sequential convex optimization algorithm that is guaranteed to converge to a feasible point, and under suitable conditions for the functions $f_i, g_i$, to a local minimum.

The CCP involves linearizing the functions $g_i$ around some reference value for the decision variables, so that at iteration $k$, we solve the convex program
\begin{subequations}
	\begin{align}
		&\min_{x\in\R^{n}} \ f_0(x) - (g_0(x_\ell) + \nabla g_0(x_\ell)^\intercal (x - x_\ell)), \\
		&\ \mathrm{s.t.} \quad f_i(x) - (g_i(x_\ell) + \nabla g_i(x_\ell)^\intercal (x - x_\ell)) \leq 0, \\
		&\qquad\qquad\qquad\qquad\qquad\qquad\qquad i = 1,\ldots, M. \nonumber
	\end{align}
\end{subequations}
For simplicity, we drop the indices $i,k$ from the chance constraints.
In the context of the DC constraints \eqref{eq:squared_CC_state1} and \eqref{eq:squared_CC_input1}, the linearizations then become
\begin{subequations}
	\begin{align*}
		g_{x}(\mu) &\approx (\beta - \alpha^\intercal \mu_\ell)^2 - 2\alpha^\intercal(\beta - \alpha^\intercal\mu_\ell)(\mu - \mu_{\ell}), \\
		g_{u}(m) &\approx  (b - a^\intercal m_{\ell})^{2} - 2a^\intercal(b - a^\intercal m_{\ell})(m - m_{\ell}).
	\end{align*}
\end{subequations}

\end{document}